\documentstyle[11pt]{article}

\newcommand{\noi}{\vspace{12pt}\noindent}
\newcommand{\beq}{\begin{equation}}
\newcommand{\eeq}{\end{equation}}
\newcommand{\bea}{\begin{eqnarray}}
\newcommand{\eea}{\end{eqnarray}}

\newcommand{\e}[1]{{(\ref{#1})}}
\newcommand{\eq}[1]{{eq.\ (\ref{#1})}}
\newcommand{\es}[2]{{(\ref{#1}) and (\ref{#2})}}
\newcommand{\eqs}[2]{{eqs.\ (\ref{#1}) and (\ref{#2})}}
\newcommand{\Ref}[1]{{Ref.~\cite{#1}}}
\newcommand{\mb}[1]{{\mbox{${#1}$}}}

\newcommand{\ie}{{${ i.e.\ }$}}
\newcommand{\eg}{{${ e.g.\ }$}}

\newcommand{\wrt}{{with respect to }}

\newcommand{\rhs}{{right-hand side }}

\newcommand{\sdet}{{\rm sdet}}

\newcommand{\Hf}{{1 \over 2}}

\newcommand{\Ih}{{i \over \hbar}}

\newcommand{\deder}[1]{{ 
 {\stackrel{\raise.1ex\hbox{$\leftarrow$}}{\delta^r}   } 
\over {   \delta {#1}}  }}
\newcommand{\dedel}[1]{{ 
 {\stackrel{\lower.3ex \hbox{$\rightarrow$}}{\delta^l}   }
 \over {   \delta {#1}}  }}
\newcommand{\papar}[1]{{ 
 {\stackrel{\raise.1ex\hbox{$\leftarrow$}}{\partial^r}   } 
\over {   \partial {#1}}  }}
\newcommand{\papal}[1]{{ 
 {\stackrel{\lower.3ex \hbox{$\rightarrow$}}{\partial^l}   }
 \over {   \partial {#1}}  }}
\newcommand{\ddr}[1]{{ 
 {\stackrel{\raise.1ex\hbox{$\leftarrow$}}{\delta^r}   } 
\over {   \delta {#1}}  }}
\newcommand{\ddl}[1]{{ 
 {\stackrel{\lower.3ex \hbox{$\rightarrow$}}{\delta^l}   }
 \over {   \delta {#1}}  }}

\newcommand{\proofbox}{\begin{flushright}
${\,\lower0.9pt\vbox{\hrule \hbox{\vrule
height 0.2 cm \hskip 0.2 cm \vrule height 0.2 cm}\hrule}\,}$
\end{flushright}}

\newtheorem{theorem}{Theorem}[section]

\newtheorem{lemma}[theorem]{Lemma}

\begin{document}
\thispagestyle{empty}
\title{\Large{\bf A Note on Semidensities in Antisymplectic Geometry}}
\author{{\sc K.~Bering}$^1$\\Institute for Theoretical Physics \& Astrophysics
\\Masaryk University\\Kotl\'a\v{r}sk\'a 2\\CZ-611 37 Brno\\Czech Republic}
\maketitle
\vfill
\begin{abstract}
We revisit Khudaverdian's geometric construction of an odd nilpotent operator
$\Delta_E$ that sends semidensities to semidensities on an antisymplectic
manifold. We find a local formula for the $\Delta_E$ operator in arbitrary
coordinates and we discuss its connection to Batalin-Vilkovisky quantization.
\end{abstract}
\vfill
\begin{quote}
MCS number(s): 53A55; 58A50; 58C50; 81T70. 
\hrule width 5.cm \vskip 2.mm \noindent 
$^{1}${\small E-mail:~{\tt bering@physics.muni.cz}} \\ 
\end{quote}

\newpage

\setcounter{equation}{0}
\section{Introduction}

\noi
Recall that for a symplectic manifold with an even symplectic two-form 
\mb{\omega=\Hf dz^{A}\omega_{AB}dz^{B}}, there exists a canonical measure
density given by the Pfaffian \mb{\rho\!=\!{\rm Pf}(\omega_{AB})}, \ie there 
is a natural notion of volume in a symplectic manifold. A related fact is the 
Liouville Theorem, which states that Hamiltonian vector fields are 
divergenceless. On the other hand, the situation is completely different for
an odd symplectic manifold, also known as an antisymplectic manifold
and endowed with an odd antisymplectic two-form 
\mb{E=\Hf d\Gamma^{A}E_{AB}d\Gamma^{B}}. These geometries for 
instance show up in the Lagrangian quantization method of Batalin and
Vilkovisky \cite{bv81}. It turns out that there is {\em no}
canonical choice of measure density \mb{\rho} in this case, as, for instance,
the above Pfaffian. This is tied to the fact that there is no meaningful
notion of a superdeterminant/Berezinian for a matrix that is intrinsically 
Grassmann-odd. However, the upset runs deeper. In fact, a density \mb{\rho}
can never be a function of the antisymplectic matrix \mb{E_{AB}}. 
Phrased differently, a density \mb{\rho} always carries information that
cannot be deduced from the antisymplectic structure \mb{E} alone \cite{b97}.
Within the standard Batalin-Vilkovisky framework, the possible choices of a
density  \mb{\rho} is only partially determined by a requirement of gauge
symmetry. 

\noi
Around 1992 Batalin-Vilkovisky quantization took a more geometric form, in
particular with the work of Schwarz \cite{schwarz93}. The concensus was that
the geometric setting requires two independent structures: an odd symplectic,
non-degenerate two-form \mb{E} and a measure density \mb{\rho}. {}From these 
two structures, one may build a 
Grassmann-odd, second-order operator \mb{\Delta_{\rho}}, known as the odd 
Laplacian. Alternatively, one can view the odd Laplacian \mb{\Delta_{\rho}}
itself as {\em the} fundamental structure of Batalin-Vilkovisky geometry 
\cite{bt93,bbd96}, which is conventionally required to be nilpotent.

\noi
Khudaverdian has constructed \cite{k99,kv02,k02,k04} a Grassmann-odd,
nilpotent, second-order operator \mb{\Delta_{E}} that does not rely on a
choice of density \mb{\rho}. The caveat is that the \mb{\Delta_{E}} operator
is defined on semidensities rather than on scalars. (The notion of a
semidensity is explained in \eq{coordtransf} below.)
In retrospect, many pieces of Khudaverdian's
construction were known to physicists, see for instance \Ref{henneaux92},
p.440. In this short note we reconsider Khudaverdian's construction and find
a local formula for the \mb{\Delta_{E}} operator that applies to arbitrary
coordinate systems. The ability to work in any coordinates, not just Darboux
coordinates, is important, since if one first has to search for a set of
Darboux coordinates to the system that one is studying, symmetries (such as,
\eg, Lorentz covariance) or locality that one would like to preserve during
the quantization process, are often lost. 
 
\noi
The paper is organized as follows: We consider the antisymplectic structure 
in Section~\ref{secbvgeometry}; the odd Laplacian \mb{\Delta_{\rho}} in
Section~\ref{secoddlapl}; and in Sections~\ref{secsemi} and \ref{secarbi},
the \mb{\Delta_{E}} operator using Darboux coordinates and general
coordinates, respectively. {}Finally, in Section~\ref{secappl} we analyze a
modified Batalin-Vilkovisky scheme based on the \mb{\Delta_{E}} operator.

\noi
{\em General remark about notation}. We have two types of grading: A Grassmann
grading \mb{\epsilon} and an exterior form degree \mb{p}. The sign conventions
are such that two exterior forms \mb{\xi} and \mb{\eta}, of Grassmann parity
\mb{\epsilon_{\xi}}, \mb{\epsilon_{\eta}} and exterior form degree
\mb{p_{\xi}}, \mb{p_{\eta}}, respectively, commute in the following graded
sense
\beq
 \eta \wedge \xi
~=~(-1)^{\epsilon_{\xi}\epsilon_{\eta}+p_{\xi}p_{\eta}}\xi\wedge \eta
\eeq
inside the exterior algebra.
We will often not write the exterior wedges ``\mb{\wedge}'' explicitly.

\setcounter{equation}{0}
\section{Antisymplectic Geometry}
\label{secbvgeometry}

\noi
Consider an antisymplectic manifold \mb{(M,E)} and let \mb{\Gamma^{A}}
denote local coordinates of Grassmann parity 
\mb{\epsilon_{A}\equiv\epsilon(\Gamma^{A})}
(and exterior form degree \mb{p(\Gamma^{A})=0}).
The antisymplectic two-form can locally be written as
\beq
E~=~\Hf d\Gamma^{A}~E_{AB}~d\Gamma^{B}
~=~-\Hf E_{AB}~d\Gamma^{B}~d\Gamma^{A}~,
\eeq
where \mb{E_{AB}\!=\!E_{AB}(\Gamma)} is the corresponding matrix 
representation. Besides carrying gradings \mb{\epsilon(E)=1} and \mb{p(E)=2},
the antisymplectic two-form \mb{E} has two defining properties. {}First, 
\mb{E} is closed,
\beq
 dE~=~0~, \label{eclosed}
\eeq
where the grading conventions for the exterior derivative
\beq
d~=~d\Gamma^{A} \papal{\Gamma^{A}}
\eeq
are \mb{\epsilon(d)=0} and \mb{p(d)=1}. Secondly, \mb{E} is non-degenerate,
\ie the antisymplectic matrix \mb{E_{AB}} has an inverse matrix \mb{E^{AB}},
\beq
 E^{AB} E_{BC}~=~\delta^{A}_{C}~=~E_{CB} E^{BA} ~.
\eeq
Instead of the compact exterior form notation \mb{E},
one may equivalently formulate the above conditions with all the indices 
written out explicitly in terms of the matrices \mb{E_{AB}} or \mb{E^{AB}}.
In detail, the gradings are
\beq
\begin{array}{rcccl}
\epsilon(E_{AB})&=&\epsilon_{A}+ \epsilon_{B}+1&=&\epsilon(E^{AB})~, \cr
p(E_{AB})&=&0&=&p(E^{AB})~,
\end{array}
\label{egrad}
\eeq
the skew-symmetries are
\bea
E_{BA}&=&-(-1)^{\epsilon_A\epsilon_B}E_{AB}~, \cr
E^{BA}&=&-(-1)^{(\epsilon_A+1)(\epsilon_B+1)} E^{AB}~,
\label{esym}
\eea
while the closeness condition and the equivalent Jacobi identity read
\bea
\sum_{{\rm cycl.}~A,B,C}(-1)^{\epsilon_A \epsilon_C}
(\papal{\Gamma^A} E_{BC} ) &=&0~, \label{eclosedabc} \\
\sum_{{\rm cycl.}~A,B,C}(-1)^{(\epsilon_A+1)( \epsilon_C+1)}
E^{AD} ( \papal{\Gamma^D} E^{BC}) &=&0~, \label{ejacid}
\eea
respectively.
The inverse matrix \mb{E^{AB}} with upper indices gives rise to the 
antibracket \cite{bv81}
\beq
(F,G)~=~(F\papar{\Gamma^{A}}) E^{AB}(\papal{\Gamma^{B}}G)~,
\label{antibracket}
\eeq
which satisfies a graded skew-symmetry and a graded Jacobi identity as a 
consequence of \eqs{esym}{ejacid}. There is an antisymplectic analogue of
Darboux's Theorem that states that locally there exist Darboux coordinates
\mb{\Gamma^{A}\!=\!\left\{\phi^{\alpha};\phi^{*}_{\alpha}\right\}}, such that
the only non-vanishing antibrackets between the coordinates are
\mb{(\phi^{\alpha},\phi^{*}_{\beta})=\delta^{\alpha}_{\beta}
=-(\phi^{*}_{\beta},\phi^{\alpha})}. In Darboux coordinates the antisymplectic
two-form is simply \mb{E=d\phi^{*}_{\alpha}\wedge d\phi^{\alpha}}.

\setcounter{equation}{0}
\section{Odd Laplacian \mb{\Delta_{\rho}} on Scalars}
\label{secoddlapl}

\noi
A scalar function \mb{F\!=\!F(\Gamma)}, a density \mb{\rho\!=\!\rho(\Gamma)} 
and a semidensity \mb{\sigma\!=\!\sigma(\Gamma)} are by definition quantities
that transform as
\beq
F~~\longrightarrow~F^{\prime}~=~F~,~~~~~~~~~~~
\rho~~\longrightarrow~~\rho^{\prime}~=~\frac{\rho}{J}~,~~~~~~~~~~~
\sigma~~\longrightarrow~~\sigma^{\prime}~=~\frac{\sigma}{\sqrt{J}}~,
\label{coordtransf}
\eeq
respectively, under general coordinate transformations
\mb{\Gamma^{A}\to\Gamma^{\prime A}}, where 
\mb{J\equiv\sdet\frac{\partial \Gamma^{\prime A}}{\partial \Gamma^{B}}}
denotes the Jacobian.  We shall ignore the global issues of orientation
and choice of square root.
In principle the above \mb{F}, \mb{\rho} and \mb{\sigma}
could either be bosons or fermions, however normally we shall require the
densities \mb{\rho} to be invertible, and therefore bosons.

\noi
Given a choice of density \mb{\rho} one may define the odd Laplacian
\cite{bt93}
\beq
\Delta_{\rho}~:=~\frac{(-1)^{\epsilon_{A}}}{2\rho}
\papal{\Gamma^{A}}\rho E^{AB}\papal{\Gamma^{B}}~,\label{deltarho}
\eeq
that takes scalars to scalars of opposite Grassmann parity.
The odd Laplacian \e{deltarho} has a geometric interpretation as a 
divergence of a Hamiltonian vector field \cite{schwarz93,kn93}
\beq
  \Delta_{\rho}\Psi~=~-\Hf {\rm div}_{\rho}(X_{\Psi})
~,~~~~~~~~~~~~~~~~~\epsilon(\Psi)~=~1~.
\label{deltadivham}
\eeq
Here \mb{X_{\Psi}:=(\Psi,\cdot)} denotes a Hamiltonian vector field with a
Grassmann-odd Hamiltonian \mb{\Psi}, and the divergence \mb{{\rm div}_{\rho}X}
of a vector field \mb{X}, \wrt the measure density \mb{\rho}, is
\beq
{\rm div}_{\rho}X~:=~
\frac{(-1)^{\epsilon_{A}}}{\rho}\papal{\Gamma^{A}}(\rho X^{A})~,~~~~~~~~~~
\epsilon(X)~=~0~.
\eeq
The fact that the odd Laplacian \e{deltadivham} is non-zero, shows that 
antisymplectic manifolds do not have an analogue of the Liouville Theorem
mentioned in the Introduction.
As a consequence of the Jacobi identity \eq{ejacid}, the square operator
\mb{\Delta_{\rho}^{2}=\Hf[\Delta_{\rho},\Delta_{\rho}]} becomes a linear
derivation, \ie a first-order differential operator,
\beq
\Delta_{\rho}^{2}(F G) ~=~\Delta_{\rho}^{2}(F)~G+F~\Delta_{\rho}^{2}(G)~.
\label{delta2leibnitz}
\eeq
Conventionally, one imposes additionally that the \mb{\Delta_{\rho}} operator
is nilpotent \mb{\Delta_{\rho}^{2}=0}, but this is not necessary for our
purposes.

\setcounter{equation}{0}
\section{Khudaverdian's \mb{\Delta_{E}} Operator on Semidensities}
\label{secsemi}

\noi
Khudaverdian showed that one may define a Grassmann-odd, nilpotent,
second-order operator \mb{\Delta_{E}} {\em without} a choice of density
\mb{\rho}. This \mb{\Delta_{E}} operator does not take scalars to scalars like
the odd Laplacian \e{deltarho}, but instead takes semidensities to 
semidensities of opposite Grassmann parity. 
Equivalently, the \mb{\Delta_{E}} operator transforms as
\beq
\Delta_{E}~~\longrightarrow~\Delta_{E}^{\prime}
~=~\frac{1}{\sqrt{J}}\Delta_{E}\sqrt{J} 
\label{deltaetransf}
\eeq
under general coordinate transformations \mb{\Gamma^{A}\to\Gamma^{\prime A}},
cf.\ \eq{coordtransf}. Khudaverdian's construction relies first of
all on an atlas of Darboux charts, which is granted by an antisymplectic
analogue of Darboux's Theorem, and secondly, on a Lemma by Batalin and
Vilkovisky about the possible form of the Jacobians for anticanonical
transformations, also known as antisymplectomorphisms. 

\begin{lemma} 
``The Batalin-Vilkovisky Lemma'' \cite{bv84,henneaux92,kv02,k02,k04,bbd06}.
Consider a finite anticanonical transformation between initial Darboux
coordinates \mb{\Gamma^{A}_{(i)}} and final Darboux coordinates 
\mb{\Gamma^{A}_{(f)}}. Then the Jacobian 
\mb{J\equiv\sdet(
\partial \Gamma^{A}_{(f)} /\partial \Gamma^{B}_{(i)})} 
satisfies
\beq
\Delta_{1}^{(i)} \sqrt{J}~=~0~.
\label{bvlemma}
\eeq
Here \mb{\Delta_{1}^{(i)}} refers to the odd Laplacian \e{deltarho} with
\mb{\rho\!=\!1} in the initial Darboux coordinates \mb{\Gamma^{A}_{(i)}}.
\label{thebvlemma}
\end{lemma}

\noi
Given Darboux coordinates \mb{\Gamma^{A}} the \mb{\Delta_{E}} operator is 
defined on a semidensity \mb{\sigma} as \cite{k99,kv02,k02,k04,bbd06}
\beq
(\Delta_{E}\sigma)~:=~(\Delta_{1}\sigma)~,\label{khudeltaesigma}
\eeq
where \mb{\Delta_{1}} is the \mb{\Delta_{\rho}} operator \e{deltarho} with
\mb{\rho\!=\!1}.  It is important in \eq{khudeltaesigma} that the formula for
the \mb{\Delta_{1}} operator \e{deltarho} and the semidensity \mb{\sigma}
both refer to the same Darboux coordinates \mb{\Gamma^{A}}. The parentheses in
\eq{khudeltaesigma} indicate that the equation should be understood as an
equality among semidensities (in the sense of zeroth-order differential
operators) rather than an identity among differential operators. One next uses
the Batalin-Vilkovisky Lemma to argue that the definition
\e{khudeltaesigma} does not depend on the choices of Darboux coordinates
\mb{\Gamma^{A}}. What this means is,  that the \rhs of the definition
\e{khudeltaesigma} transforms as a semidensity
\beq
 (\Delta^{(f)}_{1}\sigma_{(f)})
~=~\frac{1}{\sqrt{J}}(\Delta^{(i)}_{1}\sigma_{(i)})
\eeq
under an anticanonical transformation between any two Darboux coordinates 
\mb{\Gamma^{A}_{(i)}} and \mb{\Gamma^{A}_{(f)}}. Proof:
\beq
\sqrt{J}(\Delta^{(f)}_{1}\sigma_{(f)})
~=~\sqrt{J}(\Delta^{(i)}_{J}\sigma_{(f)})
~=~\sqrt{J}(\Delta^{(i)}_{J}\frac{\sigma_{(i)}}{\sqrt{J}})
~=~(\Delta^{(i)}_{1}\sigma_{(i)})
-\frac{1}{\sqrt{J}}(\Delta^{(i)}_{1}\sqrt{J})\sigma_{(i)}
~=~(\Delta^{(i)}_{1}\sigma_{(i)})~.
\eeq
The third equality is a non-trivial property of the odd Laplacian 
\e{deltarho}. The Batalin-Vilkovisky Lemma is used in the fourth equality.
Strictly speaking, it is enough to consider infinitesimal anticanonical
transformations to justify the definition \e{khudeltaesigma}. The proof of
the infinitesimal version of the Batalin-Vilkovisky Lemma goes like this: An
infinitesimal anticanonical coordinate transformation 
\mb{\delta\Gamma^{A}=X^{A}} is necessarily a Hamiltonian vector field
\mb{X^{A}=(\Psi,\Gamma^{A})\equiv X_{\Psi}^{A}} with an infinitesimal,
Grassmann-odd Hamiltonian \mb{\Psi}, where \mb{\epsilon(\Psi)=1}. So 
\beq 
\ln J~\approx ~(-1)^{\epsilon_{A}}(\papal{\Gamma^{A}} X^{A})
~=~{\rm div_{1}}(X_{\Psi})~=~-2\Delta_{1}\Psi~,
\eeq
and hence
\beq
\Delta_{1}\sqrt{J}~\approx~-\Delta^{2}_{1}\Psi~=~0~,
\eeq
due to the nilpotency of the \mb{\Delta_{1}} operator in Darboux coordinates.
The ``\mb{\approx}'' sign is used to indicate that equality only holds at the
infinitesimal level. (Here we are guilty of mixing active and passive
pictures; the active vector field is properly speaking {\em minus} \mb{X}.)
A simple proof of the Batalin-Vilkovisky Lemma for finite anticanonical
transformations can be found in \Ref{bbd06}.

\noi
On the other hand, once the definition \e{khudeltaesigma} is justified,
it is obvious that the \mb{\Delta_{E}} operator super-commutes with itself,
because the \mb{\Gamma^{A}}-derivatives have no \mb{\Gamma^{A}}'s to act on
in Darboux coordinates. Therefore \mb{\Delta_{E}} is nilpotent,
\beq
\Delta_{E}^{2}~=~\Hf [\Delta_{E},\Delta_{E}]~=~0~.\label{deltaenilp}
\eeq
Same sort of reasoning shows that \mb{\Delta_{E}=\Delta_{E}^{T}} is symmetric.

\setcounter{equation}{0}
\section{The \mb{\Delta_{E}} Operator in General Coordinates}
\label{secarbi}

\noi
We now give a definition of the \mb{\Delta_{E}} operator that does not
rely on Darboux coordinates. We claim that in arbitrary coordinates the
\mb{\Delta_{E}} operator is given as
\beq
 (\Delta_{E}\sigma)~:=~ (\Delta_{1}\sigma)
+\left(\frac{\nu^{(1)}}{8}-\frac{\nu^{(2)}}{24} \right)\sigma~,
\label{ourdeltaesigma}
\eeq
where
\bea
\nu^{(1)}&:=&
(-1)^{\epsilon_{A}}(\papal{\Gamma^{B}}\papal{\Gamma^{A}}E^{AB})~,\\
\nu^{(2)}&:=&-(-1)^{\epsilon_{B}}(\Gamma^{C},(\Gamma^{B},\Gamma^{A}))
(\papal{\Gamma^{A}}E_{BC})
~=~ (-1)^{\epsilon_{A}\epsilon_{C}}(\papal{\Gamma^{A}}E^{CD})
(\papal{\Gamma^{D}}E^{AB})E_{BC}~.
\eea
Eq.\ \e{ourdeltaesigma} is the main result of this paper. Notice that in
Darboux coordinates, where \mb{E^{AB}} is constant, \ie independent of the
coordinates \mb{\Gamma^{A}}, the last two terms \mb{\nu^{(1)}} and
\mb{\nu^{(2)}} vanish. Hence the definition \e{ourdeltaesigma} agrees in this
case with Khudaverdian's \mb{\Delta_{E}} operator \e{khudeltaesigma}.

\noi
It remains to be shown that the \rhs of \eq{ourdeltaesigma} behaves as a 
semidensity under general coordinate transforms. Here we will only explicitly
consider the case where \mb{\sigma} is invertible to simplify the
presentation. (The non-invertible case is fundamentally no different.) In the
invertible case, we customarily write the semidensity \mb{\sigma=\sqrt{\rho}}
as a square root of a density \mb{\rho}, and define a Grassmann-odd quantity
\beq
 \nu_{\rho}~:=~ \frac{1}{\sqrt{\rho}}(\Delta_{E}\sqrt{\rho})
~=~ \nu_{\rho}^{(0)}+\frac{\nu^{(1)}}{8}-\frac{\nu^{(2)}}{24}~,
\label{nurho}
\eeq
by dividing both sides of the definition \e{ourdeltaesigma} with the
semidensity \mb{\sigma}. Here we have defined
\beq
 \nu_{\rho}^{(0)}~:=~\frac{1}{\sqrt{\rho}}(\Delta_{1}\sqrt{\rho})~.
\label{nurho0}
\eeq
Hence, to justify the definition \e{ourdeltaesigma}, one should check that
\mb{\nu_{\rho}} is a scalar under general infinitesimal coordinate
transformations. Under an arbitrary infinitesimal coordinate transformation
\mb{\delta\Gamma^{A}=X^{A}}, one calculates
\bea
\delta\nu_{\rho}^{(0)}&=&-\Hf \Delta_{1}{\rm div}_{1}X~, \label{dnurho0}  \\
\delta\nu^{(1)}&=& 4 \Delta_{1}{\rm div}_{1}X 
+ (-1)^{\epsilon_{A}}(\papal{\Gamma^{C}}E^{AB})
(\papal{\Gamma^{B}}\papal{\Gamma^{A}}X^{C})~, \label{dnu1} \\
\delta\nu^{(2)}&=&3(-1)^{\epsilon_{A}}(\papal{\Gamma^{C}}E^{AB})
(\papal{\Gamma^{B}}\papal{\Gamma^{A}}X^{C})~, \label{dnu2}
\eea
cf.\ Appendices~\ref{appnurho0}--\ref{appnu2}.
One easily sees that while the three constituents \mb{\nu_{\rho}^{(0)}},
\mb{\nu^{(1)}} and \mb{\nu^{(2)}} separately have non-trivial transformation
properties, the linear combination \mb{\nu_{\rho}} in \eq{nurho} is indeed a 
scalar.

\noi
The new definition \e{ourdeltaesigma} is clearly symmetric 
\mb{\Delta_{E}=\Delta_{E}^{T}} and one may check that the nilpotency 
\e{deltaenilp} of the \mb{\Delta_{E}} operator \e{ourdeltaesigma} precisely
encodes the Jacobi identity \e{ejacid}. 
The odd Laplacian \mb{\Delta_{\rho}} can be expressed entirely by the 
\mb{\Delta_{E}} operator and a choice of density \mb{\rho},
\beq
(\Delta_{\rho}F)~=~(\Delta_{1}F)+\frac{1}{\sqrt{\rho}}(\sqrt{\rho},F)
~=~\frac{1}{\sqrt{\rho}}[\stackrel{\rightarrow}{\Delta}_{1},F]\sqrt{\rho}
~=~\frac{1}{\sqrt{\rho}}[\stackrel{\rightarrow}{\Delta}_{E},F]\sqrt{\rho}~.
\label{deltarhoviadeltae}
\eeq
Since \mb{\nu^{(2)}} depends on the antisymplectic matrix \mb{E_{AB}} with
lower indices, it is not clear how the formula \e{ourdeltaesigma} extends to
the degenerate anti-Poisson case.

\setcounter{equation}{0}
\section{Application to Batalin-Vilkovisky Quantization}
\label{secappl}

\noi
It is interesting to transcribe the Batalin-Vilkovisky quantization,
based on the odd Laplacian \mb{\Delta_{\rho}}, into a quantization scheme that
is based on the \mb{\Delta_{E}} operator, with the added benefit that no choice
of measure density \mb{\rho} is needed. Since the \mb{\Delta_{E}} operator
takes semidensities to semidensities, this suggests that the Boltzmann
factor \mb{\exp[\Ih W_{E}]} that appears in the Quantum Master Equation
\beq
\Delta_{E}\exp\left[\Ih W_{E}\right]~=~0 \label{qmew}
\eeq 
should now be a semidensity, where 
\beq
W_{E}~=~S+\sum_{n=1}^{\infty}(i\hbar)^{n} W_{n}
\eeq
denotes the quantum action. In fact, this was a common interpretation (when 
restricting to Darboux coordinates) prior to the introduction of a 
density \mb{\rho} around 1992, see for instance \Ref{henneaux92}, p.440-441.
If one only considers \mb{\hbar}-independent coordinate transformations
\mb{\Gamma^{A}\to\Gamma^{\prime A}} for simplicity, this implies that the
one-loop factor \mb{e^{- W_{1}}} is a semidensity, while the rest of the
quantum action, \ie the classical action \mb{S} and the higher loop
corrections \mb{W_{n}}, \mb{n\geq 2}, are scalars as usual. {}For instance,
the nilpotent operator \mb{F\mapsto e^{W_{1}}\Delta_{E}(e^{- W_{1}}F)} takes
scalars \mb{F} to scalars. 

\noi
At this stage it might be helpful to compare the above \mb{\Delta_{E}}
approach to the \mb{\Delta_{\rho}} formalism. To this end, fix a density
\mb{\rho}. Then one can define a bona fide scalar quantum action \mb{W_{\rho}}
as
\beq
  W_{\rho}~:=~W_{E}+(i\hbar)\ln\sqrt{\rho}~, \label{wrho}
\eeq
or equivalently,
\beq
e^{\Ih W_{E}}~=~\sqrt{\rho}~e^{\Ih W_{\rho}}~. \label{wrhoexp}
\eeq
This scalar action \mb{W_{\rho}} satisfies the Modified Quantum Master Equation
\beq
(\Delta_{\rho}+\nu_{\rho})\exp\left[\Ih W_{\rho}\right]~=~0~, \label{mqmew}
\eeq
cf.\ eq.\ \e{nurho}, \e{deltarhoviadeltae}, \es{qmew}{wrhoexp}. One may obtain
the Quantum Master Equation \mb{\Delta_{\rho}\exp[\Ih W_{\rho}]=0} by
additionally imposing the covariant condition \mb{\nu_{\rho}=0}, or
equivalently \mb{\Delta_{E}\sqrt{\rho}=0}. However this step is not necessary.

\noi
Returning now to the pure \mb{\Delta_{E}} approach with no \mb{\rho}, the 
finite \mb{\Delta_{E}}-exact transformations of the form
\beq
e^{\Ih W_{E}^{\prime}}~=~
e^{-[\stackrel{\rightarrow}{\Delta}_{E},\Psi]}e^{\Ih W_{E}}~,
\label{boltzmannexactvar}
\eeq
play an important r\^ole in taking solutions \mb{W_{E}} to the Quantum Master
Equation \e{qmew} into new solutions \mb{W_{E}^{\prime}}. It is implicitly
understood that all objects in \eq{boltzmannexactvar} refer to the same (but
arbitrary) coordinate frame. In general, \mb{\Psi} is a Grasmann-odd operator
that takes semidensities to semidensities. If \mb{\Psi} is a scalar function
(=zeroth-order operator), one derives
\beq
 W_{E}^{\prime} ~=~ e^{X_{\Psi}}W_{E} 
+(i\hbar) \frac{e^{X_{\Psi}}-1}{X_{\Psi}}\Delta_{E}\Psi~.
\label{Wexactvar}
\eeq
The formula \e{Wexactvar} is similar to the usual formula in the
\mb{\Delta_{\rho}} formalism \cite{bbd06}. One may check that \eq{Wexactvar}
is covariant \wrt general coordinate transformations.

\noi
The \mb{W}-\mb{X} formulation discussed in \Ref{bbd96} and \Ref{bbd06} carries
over with only minor modifications, since the \mb{\Delta_{E}} operator is
symmetric \mb{\Delta_{E}^{T}=\Delta_{E}}. In short, the \mb{W}-\mb{X}
formulation is a very general field-antifield formulation, based on two Master
actions, \mb{W_{E}} and \mb{X_{E}}, each satisfying a Quantum Master Equation.
At the operational level, {\em symmetric} means that the \mb{\Delta_{E}}
operator, sandwiched between two semidensities under a (path) integral sign,
may be moved from one semidensity to the other, using integration by part.
This is completely analogous to the symmetry of the odd Laplacian
\mb{\Delta_{\rho}=\Delta_{\rho}^{T}} itself. The \mb{X_{E}} quantum action is
a gauge-fixing part,
\beq
X_{E}~=~ G_{\alpha}\lambda^{\alpha} + (i\hbar)H_{E} + {\cal O}(\lambda^{*})~,
\eeq
which contains the gauge-fixing constraints \mb{G_{\alpha}} in involution,
\beq
(G_{\alpha},G_{\beta})~=~G_{\gamma} U^{\gamma}_{\alpha \beta}~.\label{ginvo}
\eeq
The gauge-fixing functions \mb{G_{\alpha}} implement a generalization of the
standard Batalin-Vilkovisky gauge-fixing procedure
\mb{\phi^{*}_{\alpha}=\partial \Psi / \partial \phi^{\alpha}}. In the simplest
cases, the gauge-fixing conditions \mb{G_{\alpha}=0} are enforced by
integration over the Lagrange multipliers \mb{\lambda^{\alpha}}. See 
\Ref{bbd06} for further details on the \mb{W}-\mb{X} formulation. The
pertinent measure density in the partition function 
\beq
{\cal Z}~=~\int \! [d\Gamma][d\lambda]~e^{\Ih (W_{E}+X_{E})}
\eeq
is now located inside the one-loop parts of the \mb{W_{E}} and the \mb{X_{E}}
actions. {}For instance, an on-shell expression for the one-loop factor
\mb{e^{-H_{E}}} is
\beq
e^{-H_{E}} ~=~\sqrt{J~\sdet(F^{\alpha},G_{\beta})}~,
\label{khudaverdianh}
\eeq
where \mb{J=\sdet(\partial \bar{\Gamma}^A / \partial \Gamma^B )}
denotes the Jacobian of the transformation \mb{\Gamma^{A}\to\bar{\Gamma}^{A}}
and \mb{\bar{\Gamma}^{A} \equiv \{F^{\alpha};G_{\alpha} \}}. The formula
\e{khudaverdianh} differs from the original square root formula
\cite{bt94,k94,bbd06} by not depending on a \mb{\rho} density, consistent with
the fact that \mb{e^{-H_{E}}} is no longer a scalar but a semidensity.
We recall here the main point that the one-loop factor \mb{e^{-H_{E}}} is
independent of the \mb{F^{\alpha}}'s and the partition function \mb{\cal Z}
is independent of the \mb{G_{\alpha}}'s in involution, cf.\ \eq{ginvo}.

\noi
To summarize, the density \mb{\rho} can altogether be avoided in the 
field-antifield formalism, at the cost of more complicated transformation
rules. We stress that the above transcription has no consequences for the
physics involved. {}For instance, the ambiguity that existed in the density
\mb{\rho} is still present in the choice of \mb{W_{E}} and \mb{X_{E}}.

\vspace{0.8cm}
\noindent
{\sc Acknowledgement:}~The Author thanks I.A.~Batalin and P.H.~Damgaard for
discussions. The Author also thanks the organizers of the workshop ``Gerbes,
Groupoids, and Quantum Field Theory, May--July 2006'' at the Erwin
Schr\"odinger Institute for warm hospitality. This work is supported by the
Ministry of Education of the Czech Republic under the project MSM 0021622409.

\appendix

\setcounter{equation}{0}
\section{Proof of \eq{dnurho0}}
\label{appnurho0}

\noi
Consider a general (not necessarily infinitesimal) coordinate transformation
\mb{\Gamma^{A}\to\Gamma^{\prime A}} between  an ``unprimed'' and an
``primed'' coordinate systems \mb{\Gamma^{A}} and \mb{\Gamma^{\prime A}},
respectively, cf.\ \eq{coordtransf}. The primed \mb{\nu_{\rho}^{(0)}}
quantity \e{nurho0} can be re-expressed with the help of the unprimed
coordinates as
\beq
\nu_{\rho^{\prime}}^{\prime(0)}
~:=~\frac{1}{\sqrt{\rho^{\prime}}}(\Delta^{\prime}_{1}\sqrt{\rho^{\prime}})
~=~\frac{1}{\sqrt{\rho^{\prime}}}(\Delta_{J}\sqrt{\rho^{\prime}})
~=~\frac{1}{\sqrt{\rho}}(\Delta_{1}\sqrt{\rho})
-\frac{1}{\sqrt{J}}(\Delta_{1}\sqrt{J})~=~\nu_{\rho}^{(0)}-\nu_{J}^{(0)}~,
\label{nurhozeroprime}
\eeq
where it is convenient (and natural) to introduce the quantity
\beq
\nu_{J}^{(0)}~:=\frac{1}{\sqrt{J}}(\Delta_{1}\sqrt{J})
\eeq
\wrt the unprimed reference system. The third equality in \eq{nurhozeroprime}
uses a non-trivial property of the odd Laplacian \e{deltarho}. In the
infinitesimal case \mb{\delta\Gamma^{A}=X^{A}}, the expression for the
Jacobian \mb{J} reduces to a divergence \mb{\ln J \approx {\rm div}_{1}X},
and one calculates
\beq
\delta\nu_{\rho}^{(0)}~=~\nu_{\rho^{\prime}}^{\prime(0)}-\nu_{\rho}^{(0)}
~=~ -\nu_{J}^{(0)}~~=~-\Delta_{1}(\ln \sqrt{J})-\Hf(\ln\sqrt{J},\ln\sqrt{J})
~\approx~-\Hf \Delta_{1}{\rm div}_{1}X~,
\eeq
which is \eq{dnurho0}.

\setcounter{equation}{0}
\section{Proof of \eq{dnu1}}
\label{appnu1}

\noi
The infinitesimal variation of \mb{\nu^{(1)}} yields \mb{4} contributions
to linear order in the variation \mb{\delta\Gamma^{A}=X^{A}},
\beq
 \delta\nu^{(1)}~=~- \delta\nu_{I}^{(1)}-\delta\nu_{II}^{(1)}
+\delta\nu_{III}^{(1)}+\delta\nu_{IV}^{(1)}~.
\eeq
They are
\bea
\delta\nu_{I}^{(1)}&:=&(-1)^{\epsilon_{A}} (\papal{\Gamma^{B}}X^{C})
(\papal{\Gamma^{C}}\papal{\Gamma^{A}}E^{AB})~, \\
\delta\nu_{II}^{(1)}&:=&(-1)^{\epsilon_{A}}\papal{\Gamma^{B}}\left(
(\papal{\Gamma^{A}}X^{C})(\papal{\Gamma^{C}}E^{AB})\right) 
~=~\delta\nu_{I}^{(1)}+\delta\nu_{V}^{(1)}~, \\
\delta\nu_{III}^{(1)}&:=&(-1)^{\epsilon_{A}}\papal{\Gamma^{B}}
\papal{\Gamma^{A}}\left((X^{A}\papar{\Gamma^{C}})E^{CB}\right)
~=~\delta\nu_{IV}^{(1)}~, \\
\delta\nu_{IV}^{(1)}&:=&(-1)^{\epsilon_{A}}\papal{\Gamma^{B}}
\papal{\Gamma^{A}}\left(E^{AC}(\papal{\Gamma^{C}}X^{B})\right) 
~=~\delta\nu_{I}^{(1)}+\delta\nu_{V}^{(1)} +\delta\nu_{VI}^{(1)}~, \\
\delta\nu_{V}^{(1)}&:=&(-1)^{\epsilon_{A}}(\papal{\Gamma^{C}}E^{AB})
(\papal{\Gamma^{B}}\papal{\Gamma^{A}}X^{C})~, \\
\delta\nu_{VI}^{(1)}&:=&(-1)^{\epsilon_{A}}\papal{\Gamma^{A}}\left(E^{AC}
\papal{\Gamma^{C}}\papal{\Gamma^{B}}X^{B}(-1)^{\epsilon_{B}}\right)
~=~ 2 \Delta_{1}{\rm div}_{1}X~,
\eea
where we have noted various relations among the contributions. Altogether,
the infinitesimal variation of \mb{\nu^{(1)}} becomes
\beq 
\delta\nu^{(1)}~=~\delta\nu_{V}^{(1)}+2\delta\nu_{VI}^{(1)}~,
\eeq
which is \eq{dnu1}.

\setcounter{equation}{0}
\section{Proof of \eq{dnu2}}
\label{appnu2}

\noi
The infinitesimal variation of
\beq
\nu^{(2)}~:=~ (-1)^{\epsilon_{A}\epsilon_{C}}
(\papal{\Gamma^{D}}E^{AB})E_{BC}(\papal{\Gamma^{A}}E^{CD})
\label{dnu2again}
\eeq 
yields \mb{8} contributions to linear order in the variation 
\mb{\delta\Gamma^{A}=X^{A}}, which may be organized as \mb{2\times 4} terms
\beq
\delta\nu^{(2)}~=~2(-\delta\nu_{I}^{(2)}-\delta\nu_{II}^{(2)}
+\delta\nu_{III}^{(2)}+\delta\nu_{IV}^{(2)})~,
\eeq
due to a \mb{(A,B) \leftrightarrow (D,C)} symmetry in \eq{dnu2again}. They are
\bea
\delta\nu_{I}^{(2)}&:=&(-1)^{\epsilon_{A}\epsilon_{C}}(\papal{\Gamma^{D}}
E^{AB})E_{BF}(X^{F}\papar{\Gamma^{C}})(\papal{\Gamma^{A}}E^{CD})~, \\
\delta\nu_{II}^{(2)}&:=&(-1)^{\epsilon_{A}\epsilon_{C}}(\papal{\Gamma^{D}}
E^{AB})E_{BC}(\papal{\Gamma^{A}}X^{F})(\papal{\Gamma^{F}}E^{CD})~, \\
\delta\nu_{III}^{(2)}&:=&(-1)^{\epsilon_{A}\epsilon_{C}}(\papal{\Gamma^{D}}
E^{AB})E_{BC}\papal{\Gamma^{A}}\left((X^{C}\papar{\Gamma^{F}})E^{FD}\right)
~=~\delta\nu_{I}^{(2)}+\delta\nu_{V}^{(2)}~, \\
\delta\nu_{IV}^{(2)}&:=&(-1)^{\epsilon_{A}\epsilon_{C}}(\papal{\Gamma^{D}}
E^{AB})E_{BC}\papal{\Gamma^{A}}\left(E^{CF} (\papal{\Gamma^{F}}X^{D})\right)
~=~\delta\nu_{II}^{(2)}+\delta\nu_{VI}^{(2)}~, \\
\delta\nu_{V}^{(2)}&:=&(-1)^{\epsilon_{A}\epsilon_{C}}E^{FD}
(\papal{\Gamma^{D}}E^{AB})E_{BC}(\papal{\Gamma^{A}}X^{C}\papar{\Gamma^{F}})
~=~-\delta\nu_{V}^{(2)}+\delta\nu_{VI}^{(2)}~,\label{dnu2v} \\
\delta\nu_{VI}^{(2)}&:=&(-1)^{\epsilon_{A}}(\papal{\Gamma^{C}}E^{AB})
(\papal{\Gamma^{B}}\papal{\Gamma^{A}}X^{C})~,
\eea
where we have noted various relations among the contributions. The Jacobi
identity \e{ejacid} for \mb{E^{AB}} is used in the second equality of
\eq{dnu2v}. Altogether, the infinitesimal variation of \mb{\nu^{(2)}}
becomes
\beq 
\delta\nu^{(2)}~=~3\delta\nu_{VI}^{(2)}~,
\eeq
which is \eq{dnu2}.

\end{document}